\documentstyle[11pt,newpasp,twoside,epsf]{article}
\def\mpc {\,$h^{-1}$\,Mpc}
\def\bj {b_{\rm J}}
\def\ol {\Omega_{\Lambda}}
\def\om {\Omega_{\rm M}}

\def\edcomment#1{\iffalse\marginpar{\raggedright\sl#1\/}\else\relax\fi}
\reversemarginpar

\begin{document}
\title{AGN Physics from QSO Clustering}
\author{Scott M. Croom \& Brian J. Boyle}
\affil{Anglo-Australian Observatory, PO Box 296, Epping, NSW 1710, 
Australia}
\author{Tom Shanks \& Phil Outram}
\affil{Dept. of Physics, University of Durham, South Road, 
Durham, UK}
\author{Robert J. Smith}
\affil{Astrophysics Research Institute, Liverpool John Moores
        University, Twelve Quays House, Egerton Wharf, Birkenhead,
        UK}
\author{Lance Miller}
\affil{Department of Physics, Oxford University, 1 Keble Road, Oxford, UK}
\author{Nicola Loaring}
\affil{Mullard Space Science Laboratory, Dorking, Surrey, UK}
\author{Suzanne Kenyon \& Warrick Couch}
\affil{School of Physics, University of New South Wales, Sydney, Australia}

\begin{abstract}
We review the current status of QSO clustering measurements,
particular with respect to their relevance in understanding AGN
physics.  Measurements based on the 2dF QSO Redshift Survey (2QZ) find
a scale length for QSO clustering of $s_0=5.76^{+0.17}_{-0.27}$\mpc\
at a redshift $\bar{z}\simeq1.5$, very  similar to low redshift
galaxies.  There is no evidence of evolution in the clustering of QSOs
from $z\sim0.5$ to $z\sim2.2$.  This lack of evolution and low
clustering amplitude suggests a short life time for AGN activity of
the order $\sim10^6-10^7$ years.  Large surveys such at the 2QZ and
SDSS also allow the the study of QSO environments in 3D for the first
time (at least at low redshift), early results from this work seem to
show no difference between the environments of QSOs and normal
galaxies.  Future studies e.g. measuring clustering as a function of
black hole mass, and deep QSO surveys should provide further insight
into the formation and evolution of AGN.
\end{abstract}

\section{Introduction}

Analysis of the space distribution of QSOs provides an important test
of models of AGN formation.  Large surveys such at the 2dF QSO
Redshift Survey (2QZ; Croom et al. 2001) and SDSS (Schneider et
al. 2002) provide the first opportunity to make accurate measurements
of the clustering properties of AGN.  Previous samples were too small
and/or inhomogeneous to make anything other than a detection of the
clustering signal (e.g. La Franca et al. 1998; Croom et al. 1996;
Shanks et al. 1987; Osmer 1981).  However, with homogeneous samples in
excess of 20000 objects, the predictions of QSO formation models can
be directly tested.  Because of the high redshift and large volume
that QSOs can sample, they are also a powerful probe of large-scale
structure which can be used to answer cosmological questions, such as
the values of fundamental cosmological parameters.  However in this
review we will concentrate on what we can learn about the physics of
AGN from clustering measurements.

Under the standard paradigms of structure formation, the strength of
QSO clustering is directly related to the mass of the dark matter
halo in which the QSOs reside.  At a given redshift, the most massive
dark matter halos will be clustered more strongly than less massive
halos.  Thus QSO clustering measurements should enable us to determine
in a statistical sense the mass of the dark matter halo containing
QSOs and their host galaxies.  For a given underlying matter power
spectrum the expected number density of dark matter halos can be
derived (Press \& Schechter 1974), and then comparisons to the number
density of QSOs (the QSO luminosity function) can be used to determine
the fraction of active galaxies at any given time, and hence the
typical lifetime of activity.

\begin{figure}
\plotone{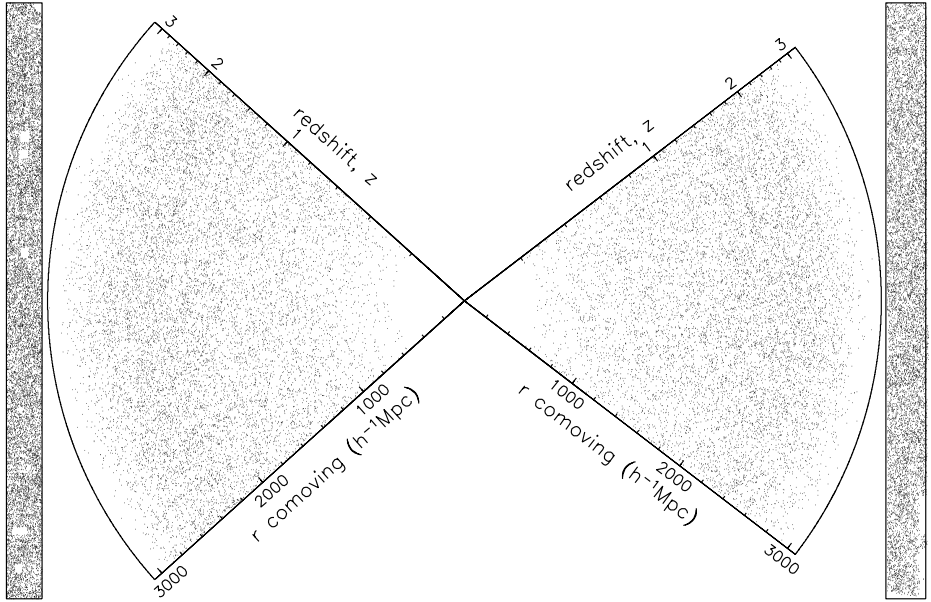}
\caption{The distribution of QSOs in the final 2QZ catalogue, showing
  the Southern (left) and equatorial (right) strips.  The rectangular
  regions show the projection onto  the sky.  An Einstein-de Sitter
  cosmology was assumed in calculating the comoving distances to the
  QSOs.}
\end{figure}

Cross-correlation with other sources (e.g. 'normal' galaxies) also
allows us to discover something about the environments of QSOs.  This
could lead to a clearer understanding of the triggering mechanisms for
activity (e.g. mergers).  

In this short review we will look at the current status of QSO
clustering measurements and QSO environmental studies.  We will
conclude with a discussion of some of the outstanding issues and
potential ways forwards.

\section{The current status of QSO clustering}

To date the most detailed clustering analysis of QSOs has been carried
out using the 2QZ sample which contains over 20000 QSOs.  Compiled
using the 2-degree Field (2dF) instrument on the Anglo-Australian Telescope
(Lewis et al. 2002), this survey obtained redshifts for QSOs at $z<3$
with $16<\bj\leq20.85$ mag.  The relatively faint magnitude limit of
the 2QZ means that the surface density of QSOs found is $\sim 35$
deg$^{-2}$, much higher than most other QSO surveys.  This high
density of QSOs makes the 2QZ a powerful probe of large-scale
structure.  The spatial distribution of QSOs in the 2QZ is shown in
Fig. 1.

The most basic of clustering measurements is the two-point correlation
function, $\xi(s)$ (or its Fourier transform the power spectrum,
$P(k)$), where $s$ is the separation of QSOs in redshift-space.  This
is shown in Fig. 2 (left) for the 2QZ averaged over the redshift range
$0.3<z<2.9$, assuming $\om=0.3$ and $\ol=0.7$ (which we will call the
$\Lambda$ cosmology).  The measured $\xi(s)$ is very similar to that
found for low redshift galaxy samples, (e.g. the 2dF Galaxy
Redshift Survey, 2dFGRS; Hawkins et al. 2003) and the best fit power
law is $\xi(s)=(s/5.76_{-0.27}^{+0.17})^{-1.64_{-0.03}^{+0.06}}$.  We can
also fit more physically motivated models, such as those based on CDM.
In Fig. 2 we plot a linear CDM correlation function normalized such
that normal galaxies are virtually unbiased at the present day (see
Hawkins et al. 2003) (lower dotted line).  The best fit CDM models
have a shape parameter $\Gamma=0.1$, with slightly more large-scale
structure than standard models (but not significantly so).  We also
derive the non-linear corrections to these linear models based on the
formalism of Peacock \& Dodds (1996).  It is worth noting that at the
high redshifts probed by QSOs ($\bar{z}\simeq1.5$) clustering remains
linear even at relatively small scales, $\sim5$\mpc.  For a given
cosmology, we can then derive the mean bias of the QSOs, $b_{\rm
QSO}\simeq2$ at $\bar{z}\simeq1.5$.

In Fig. 2 (right) we show the best fit clustering scale length $s_0$
(assuming a power law fit) as a function of redshift.  We find that
there is no significant evolution of clustering with redshift.  The
clustering amplitude of the QSOs is also consistent with that found in
$z\sim3$ Lyman-break galaxies (Adelberger et al. 1998).  The data are
also clearly inconsistent with linear evolution (solid line), and thus
$b_{\rm QSO}$ must be a function of redshift.  A simple model would be
to assume that QSOs were cosmologically long lived objects with ages
of order a Hubble time.  In this case their bias would evolve as
$b(z)=1+(b(0)-1)G(\om,\ol,z)$ where $b(0)$ is the bias at $z=0$ and
$G(\om,\ol,z)$ is the linear growth rate of density perturbations.  In
this model the QSOs would be formed at some arbitrarily high redshift
and move in the gravitational potential of the mass distribution.
This model (dotted line) is ruled out at high ($>99.99$ \%)
significance, demonstrating that QSOs must be short lived compared to
the age of the Universe.  A number of authors (e.g. Martini \&
Weinberg 2001) have constructed more detailed models based on the
Press-Schechter formalism to constrain the typical lifetime of QSOs
via clustering measurements.  Comparison to the current data suggest
that QSO lifetimes, at least at $z\sim2$ are $\sim10^6$ years with
typical halo masses of $\sim10^{12}M_{\sun}$.  Models which include
the effects of gas and stars (e.g. see Di Matteo et al., this volume)
increase the typical time scale, but only to $\sim10^7$ years.

\begin{figure}
\plottwo{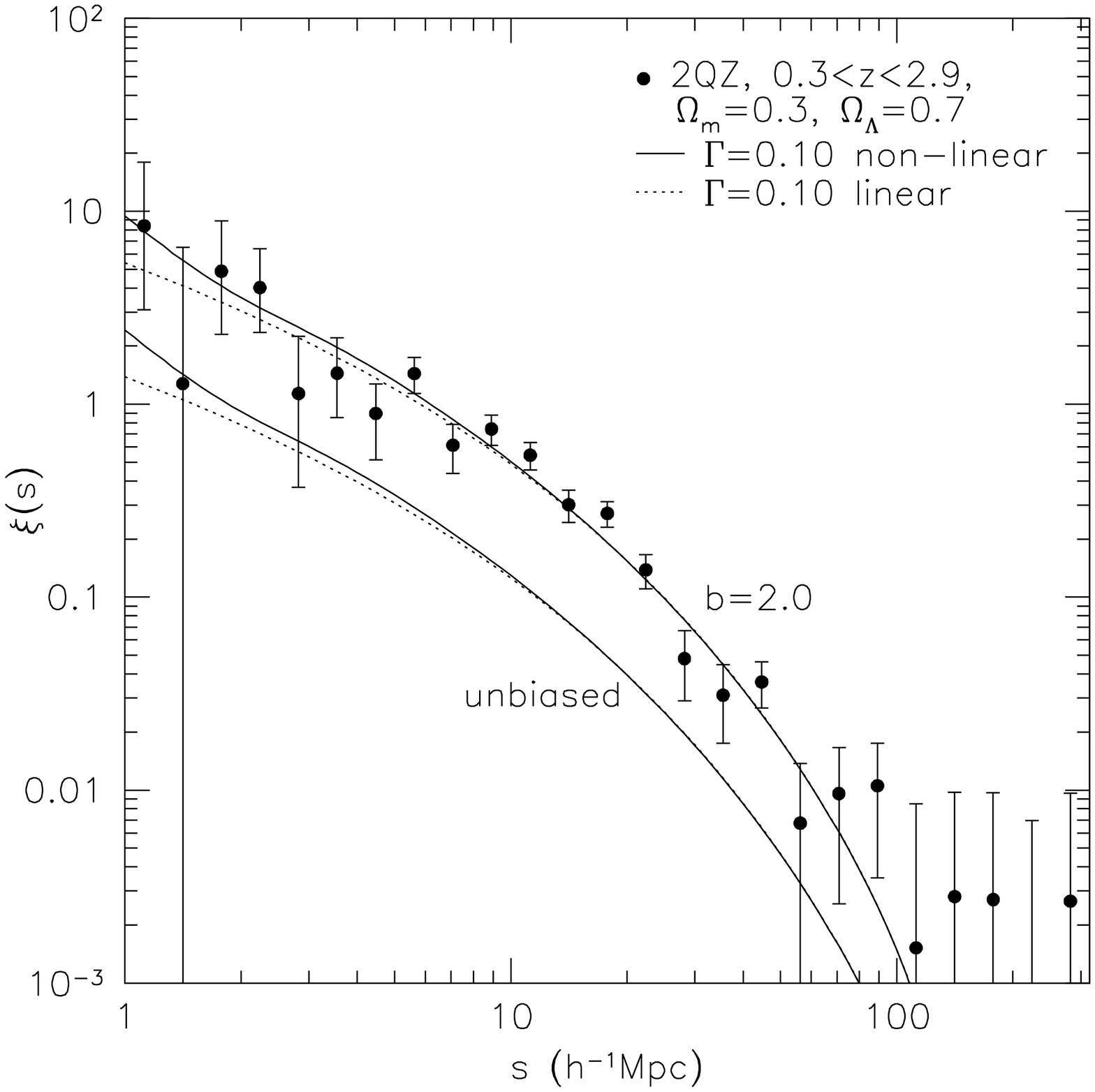}{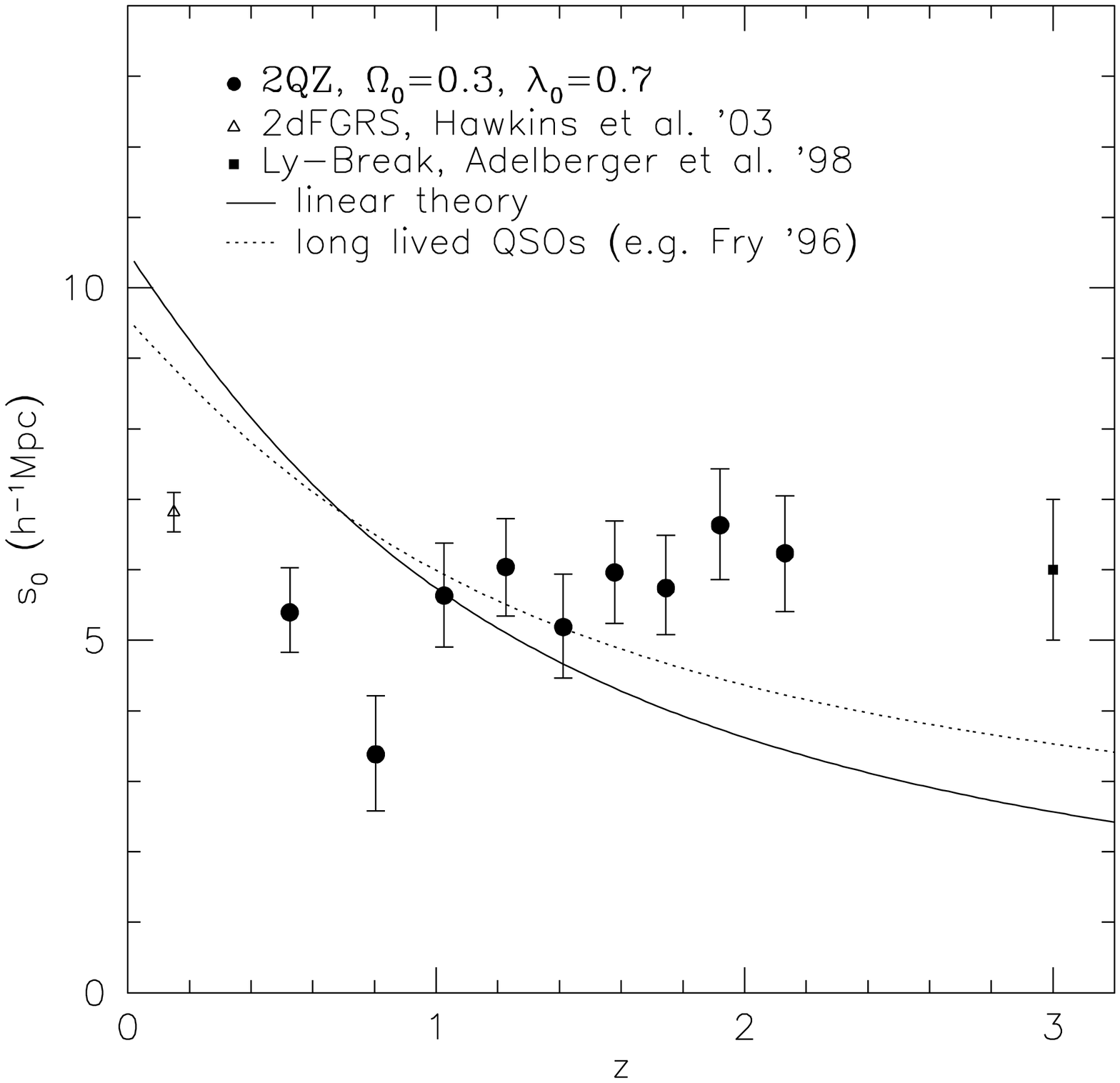}
\caption{Left: $\xi(s)$ for QSOs from the 2QZ (filled circles) in a
  $\Lambda$ Universe.  Compared to this are linear (dotted lines) and
  non-linear (solid lines) CDM models with a shape parameter,
  $\Gamma=0.1$.  The lower lines are the mass correlation function
  normalized at low redshift by observations of the 2dFGRS.  The upper
  lines are scaled by a factor of 4 ($b_{\rm QSO}=2$.).  Right:  The
  scale length, $s_0$, of QSO clustering as a function of redshift
  (filled circles) compared to the clustering of low (triangle) and
  high (square) redshift galaxies and two models assuming linear
  theory (solid line) or long lived QSOs (dotted line).}
\end{figure}

Another prediction of QSO formation models (e.g. Kauffmann \& Haehnelt
2000) is that QSO clustering depends on luminosity.  Only very
marginal evidence has so far been found to support this (e.g. Croom et
al. 2002), to fully test this prediction samples which span a broad
range in luminosity at a given redshift are required.

\section{QSO environments}

\begin{figure}
\plottwo{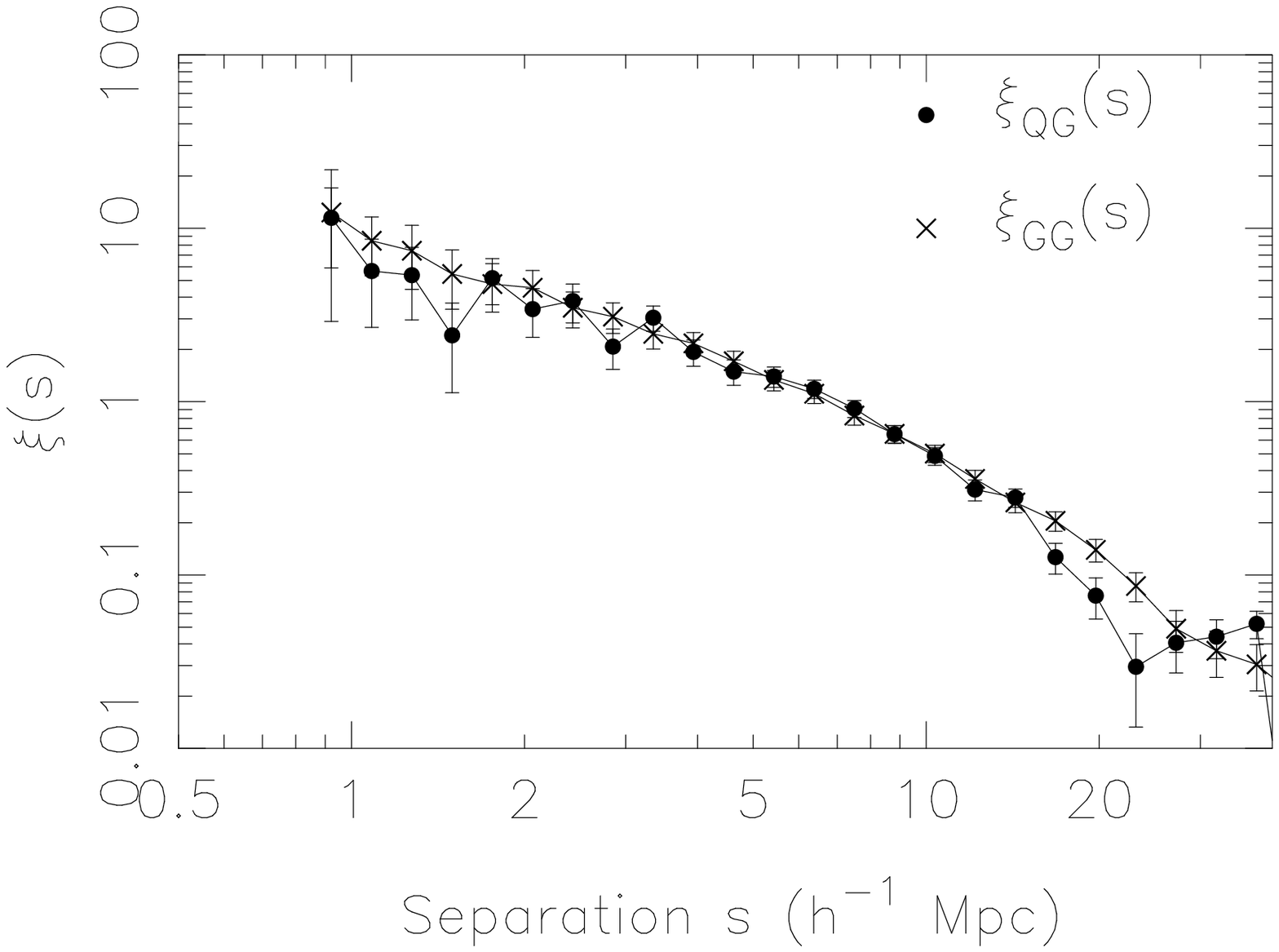}{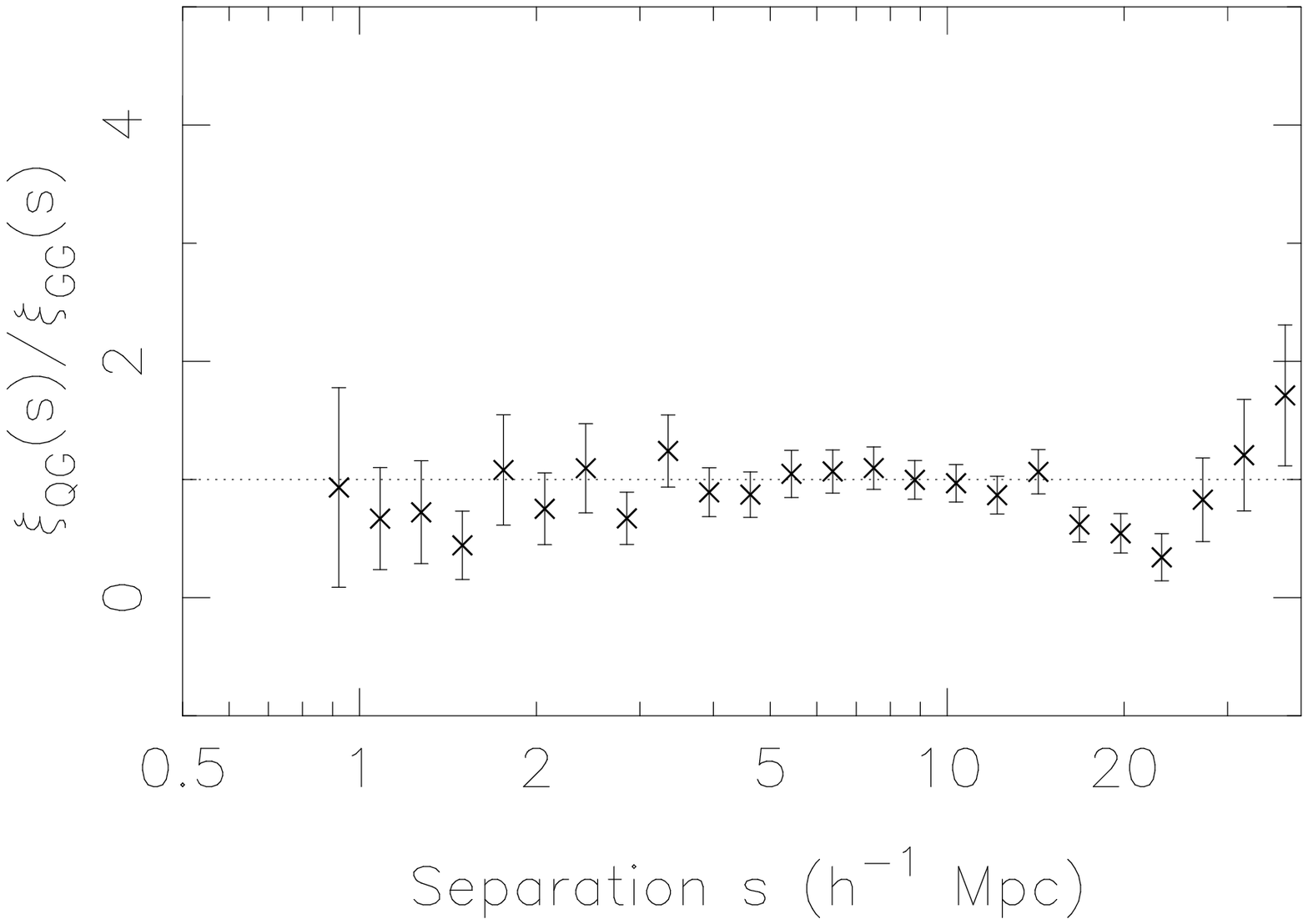}
\caption{Left: The cross-correlation, $\xi_{\rm QG}$, of 2QZ QSOs at
  $z<0.3$ with galaxies from the 2dFGRS (filled points) compared to
  the auto-correlation, $\xi_{\rm GG}$, of galaxies from the
  2dFGRS. Right: the ratio of the two, $\xi_{\rm QG}/\xi_{\rm GG}$, as
  a function of scale.}
\end{figure}

Complementary to the above analysis is the study of QSO environments
via the cross-correlation of QSOs and galaxies.  Until very recently,
this has only been possible using 2D angular clustering measurements
(e.g. Ellingson, Yee \& Green 1991; Smith, Boyle \&
Maddox 1995; Croom \& Shanks 1999).  These generally showed that
radio-quiet QSOs where clustered similarly to normal galaxies, while
radio-loud QSOs appeared in richer environments.

With new large QSO surveys such as the 2QZ and SDSS it is now possible
to carry out cross-correlations in 3D, making use of the fact that
major galaxy surveys are being carried out in the same regions of the
sky as the QSO surveys.  The cross-correlations are limited to
relatively low redshift, as these galaxy surveys typically probe
$z<0.3$, however the high quality photometry of SDSS should allow this
to be extended to higher redshift via photometric redshifts.

We have measured the cross-correlation of QSOs from the 2QZ and
galaxies in the 2dFGRS (Colless et al. 2001).  This was carried out at
$z<0.3$ where galaxies are detected.  This redshift also picks out the
intrinsically faintest QSOs which will tend to lie at low $z$ in any
flux limited sample.  The results are shown in Fig. 3.  The
cross-correlation between QSOs and galaxies is found to be identical
to the auto-correlation of galaxies (weighted to have the same
redshift distribution at the QSOs).  If we divide the two measurements
(Fig. 3 right) we determine a mean bias $b_{\rm QG}=0.97\pm0.05$.
Thus it appears that the AGN are not clustered any differently to
galaxies.  Further work is clearly required to determine whether the
bias between QSOs and galaxies is a function of QSO and/or galaxy
luminosity, and this will likely require extending the analysis to
higher redshift.  It is worth noting that this result does not rule
out the merger hypothesis for the triggering of QSOs as Percival et
al. (2003) have shown that halo merger sites are clustered the same as
non-merger halos in N-body simulations.  It is also possible to use
objects from the SDSS and 2dFGRS galaxy surveys that are shown from
their spectra to be AGN, as demonstrated recently by Miller et al
(2003).

\section{Discussion and future directions}

QSO clustering measurements are starting to place interesting
constraints on models of QSO formation.  In particular it appears that
lifetimes must be short, $\sim10^6-10^7$ years.  It is worth noting
that the clustering measurements are in fact being used as a surrogate
for host galaxy mass.  In high redshift QSOs it is very difficult to
make good estimates of galaxy mass, simply because the QSO luminosity
dwarfs the host galaxy.  If the observed relation between galaxy mass
(or more exactly galaxy velocity dispersion) and black hole mass
($M_{\rm BH}$) at low redshift (e.g. Magorrian et al. 1998) is the
same at high redshift then clustering can also be used to determine
the mean $M_{\rm BH}$ for a population.  However, it is likely that
the local galaxy-$M_{\rm BH}$ relation does evolve with redshift.
Clustering gives one potential method to determine if this is the
case.  The SDSS spectra are of sufficient quality that they should
yield reasonable estimates of $M_{\rm BH}$, thus allowing us to
measure clustering as a function of $M_{\rm BH}$.

New surveys are also required to break the still apparent
luminosity-redshift degeneracy in QSO samples.  We have very little
knowledge of QSOs more than $\sim1$ mag fainter than $M^*$,
particularly at high redshift.  A new survey based on SDSS imaging and
2dF spectroscopy is currently underway to address this issue and
others, reaching a limiting magnitude of $g'\simeq22$ for $\sim10000$
QSOs.  This is being carried out in tandem with a search for luminous
red galaxies at $z<0.7$, which will allow the investigation of QSO
environments in 3D to much higher redshifts than currently possible.
This, in combination with the high quality spectral information
available from the SDSS should allow major progress in our
understanding of QSO formation and evolution in the near future.


\begin{references}

\reference Adelberger, K. L., Steidel, C. C., Giavalisco, M.,
 Dickinson, M., Pettini, M., Kellogg, M. 1998, \apj, 505, 18

\reference Colless, M., et al. 2001, \mnras, 328, 1039

\reference Croom, S. M., Boyle, B. J., Loaring, N. S.,
Miller, L., Outram, P. J., Shanks, T., Smith, R. J. 2002, \mnras, 335,
459

\reference Croom, S.M., \& Shanks, T. 1996, \mnras, 281, 893

\reference Croom, S.M., \& Shanks, T. 1999, \mnras, 303, 411

\reference Croom, S. M., Smith, R. J., Boyle, B. J.,  Shanks, T.,
Loaring, N. S.,  Miller, L., \& Lewis I. J. 2001, \mnras, 322, L29 

\reference Ellingson, E., Yee, H. K. C., \& Green, R. F. 1991, \apj, 303, 49 

\reference Hawkins, E. et al. 2003, \mnras\ in press (astro-ph/0212375)

\reference Kauffmann, G., \& Haehnelt, M. 2000, \mnras, 311, 576

\reference La Franca, F., Andreani P., \& Cristiani, S. 1998, \apj, 497,
529 

\reference Lewis,  I. J. et al. 2002, \mnras, 333, 279

\reference Magorrian, J., et al. 1998, \aj, 115, 2285

\reference Martini, P., \& Weinberg, D. H. 2001, \apj, 547, 12

\reference Miller, C. J., Nichol, R. C., Gomez, P., Hopkins, A., \&
Bernardi, M., 2003, \apj\ in press (astro-ph/0307124)
 
\reference Osmer, P. S. 1981, \apj, 247, 762

\reference Peacock, J. A., Dodds, S. J., 1996, \mnras, 280, L19

\reference Percival, W. J., Scott, D., Peacock, J. A., \& Dunlop,
J. 2003, \mnras, 338, L31

\reference Press, W. H., Schechter, P. 1974, \apj, 187, 425

\reference Schneider, D. P., et al. 2002, \aj, 123, 567 

\reference Shanks, T., Fong, R., Boyle. B. J., \& Peterson, B. A. 1987,
\mnras, 227, 739 

\reference Smith, R. J., Boyle, B. J., \& Maddox, S. J. 1995, \mnras, 303, 270 

\end{references}
\end{document}